\documentclass[prb,floatfix,twocolumn,showpacs,amsmath,amssymb]{revtex4}
\usepackage{graphicx}% Include figure files
\usepackage{epstopdf}
\usepackage{epsfig}
\usepackage{dcolumn}% Align table columns on decimal point
\usepackage{bm}% bold math
\usepackage{amsbsy}
\usepackage[colorlinks=true,linkcolor=blue,citecolor=blue]{hyperref}

\begin{document}
\title{Variational Monte Carlo study of chiral spin liquid in quantum antiferromagnet on the triangular lattice}
\author{Wen-Jun Hu$^{1,3}$, Shou-Shu Gong$^{2,3}$, and D. N. Sheng$^3$}
\affiliation{
$^1$ Department of Physics and Astronomy, Rice University, Houston, Texas 77005, USA\\
$^2$ National High Magnetic Field Laboratory, Florida State University, Tallahassee, Florida 32310, USA\\
$^3$ Department of Physics and Astronomy, California State University, Northridge, California 91330, USA
}

\begin{abstract}
By using Gutzwiller projected fermionic wave functions and variational Monte Carlo technique,
we study the spin-$1/2$ Heisenberg model with the first-neighbor ($J_1$), second-neighbor ($J_2$),
and additional scalar chiral interaction $J_{\chi}{\bf S}_i \cdot ({\bf S}_j \times {\bf S}_k)$ on 
the triangular lattice. In the non-magnetic phase of the $J_1-J_2$ triangular model with 
$0.08 \lesssim J_2/J_1 \lesssim 0.16$, recent density-matrix renormalization group (DMRG) studies
[Zhu and White, Phys. Rev. B {\bf 92}, 041105 (2015); Hu, Gong, Zhu, and Sheng, Phys. Rev. B {\bf 92}, 140403 (2015)]
find a possible gapped spin liquid with the signal of a competition between a chiral and a $Z_2$ spin liquid. 
Motivated by the DMRG results, we consider the chiral interaction $J_{\chi}{\bf S}_i \cdot ({\bf S}_j \times {\bf S}_k)$
as a pertubation for this non-magnetic phase. We find that with growing $J_{\chi}$, 
the gapless U(1) Dirac spin liquid, which has the best variational energy for $J_{\chi}=0$, 
exhibits the energy instability towards a gapped spin liquid with non-trivial magnetic fluxes and 
nonzero chiral order. We calculate topological 
Chern number and ground-state degeneracy, both of which identify this flux state as the chiral 
spin liquid with fractionalized Chern number $C=1/2$ and two-fold topological degeneracy. Our
results indicate a positive direction to stabilize a chiral spin liquid near the non-magnetic phase of
the $J_1-J_2$ triangular model.
\end{abstract}


\maketitle

%%%%%%%%%%%%%%%%%%%%%%%%%%%%%%%%%%%%%%%%%%%%%%%%
\section{INTRODUCTION}
Quantum spin liquid is one kind of long-range entangled states
without breaking neither spin rotational nor lattice translational
symmetries even at zero temperature \cite{Balents2010, savary2016}. 
The physics of spin liquid has been playing an essential role 
to understand strongly correlated systems and unconventional 
superconductivity \cite{Anderson1973, Lee2006}. The emergent 
topological order \cite{Wen1989, WenNiu1990, wen1990topological}
and fractionalized quasiparticles \cite{Wen1991, senthil2000, senthil2001}
of spin liquid have wide applications on quantum computations and quantum communications \cite{Kitaev2006}. 
In experiment, one of the best candidates to realize spin liquid 
is frustrated antiferromagnetic material. A natural way to form geometric
frustration is to have the corner-sharing triangle and
the face-sharing triangle structures on lattice.

The simplest lattice which is constructed from corner-sharing triangles
is the kagom\'{e} lattice. At experimental side, the most promising materials
to realize spin liquid on kagom\'{e} lattice are the spin-$1/2$ antiferromagnets herbertsmithite 
and kapellasite \cite{mendels2007, helton2007, vries2009, fak2012, han2012, fu2015}. 
Theoretically, density-matrix renormalization group (DMRG) studies
consistently find a gapped spin liquid in the spin-$1/2$ kagom\'{e} Heisenberg
model with the nearest-neighbor (NN) interactions \cite{Yan2011,Depenbrock2012, Jiang2012nature}.
However, the variational studies based on projected fermionic parton wave functions favor a gapless 
U(1) Dirac spin liquid (DSL) with competing ground-state energy \cite{Ran2007,Iqbal2011,Iqbal2013,Iqbal2014}.
Near the NN model, a robust chiral spin liquid (CSL) \cite{kalmeyer1987,wen1989csl, kun1993, Greiter2007} is unambiguously
established by introducing second- and third-neighbor couplings or chiral interaction
\cite{messio2012,gong2014kagome,he2014csl,Ssgong2015,bauer2014,hu2015csl,lauchli2015}.
This CSL spontaneously breaks time-reversal symmetry (TRS) and is identified as
the $\nu = 1/2$ bosonic fractional quantum Hall state.

On the other hand, the typical system with face-sharing triangles is the simple spin-$1/2$
triangular lattice system. The NN Heisenberg antiferromagnetic model on the triangular lattice is the first candidate proposed to
realize a spin liquid \cite{Anderson1973}; however, a $120^{\circ}$ antiferromagnetic order is found
in the subsequent studies \cite{sachdev1992, bernu1992, sorella1999, zheng2006, white2007}.
Although spin liquid does not exist in the NN model, both experimental and theoretical studies find that the
additional interactions may open a new route for realizing such states. In the organic weak 
Mott insulators with triangular lattice structure such as $\kappa$-(ET)$_2$Cu$_2$(CN)$_3$
 and EtMe$_3$Sb[Pd(dmit)$_2$]$_2$ \cite{shimizu2003, kurosaki2005, yamashita2008, yamashita2009,itou2008, yamashita2010},
no magnetic order is observed at the temperature much lower
than the interaction energy scale. The spin liquid behaviors are explained by a
gapless spin Bose metal state realized in a triangular model with four-site ring-exchange interactions
\cite{lesik2005, sheng2009, matthew2011, Misguich1999}.
The spatial anisotropic triangular model with the NN couplings $J_1-J^{\prime}_1$ 
has also been studied extensively to find a possible spin liquid state at the neighbor of 
the spin spiral phase~\cite{yunoki2006, sheng2006, leon2007, white2011, federico2016}.
Recently, different theoretical studies  consistently find a non-magnetic phase in the 
$J_1-J_2$ triangular Heisenberg model, which is sandwiched between the $120^{\circ}$ and the stripe 
magnetic order phases for $0.08 \lesssim J_2/J_1 \lesssim 0.15$ 
\cite{manuel1999, mishmash2013, kaneko2014, campbell2015, zhuzhenyue2015, Hu2015, 
saadatmand2015, bishop2015, Iqbal2016}.
DMRG results suggest a gapped spin liquid for this non-magnetic phase \cite{zhuzhenyue2015, Hu2015}.
However, the finite-size DMRG calculations find numerical signals for both
CSL and $Z_2$ spin liquid~\cite{Hu2015}, which may imply strong finite-size effects;
therefore, the system has difficulty to settle into one state. On the other hand, recent variational Monte Carlo
studies \cite{Iqbal2016} find that the gapless U(1) DSL hosts the best variational energy
than the various $Z_2$ spin liquids in parton constructions \cite{QiY2015, Lu2015}.
Now, the understanding  of this non-magnetic phase in triangular model is 
in the similar situation as the NN kagom\'e model,
both of which exhibit various candidate ground states with
close energies. Inspired by the CSL signals in the triangular model \cite{Hu2015} and
the emerging CSL in kagom\'e model by considering different pertubations,
we address the issue that whether a CSL might also be stabilized by introducing further 
perturbations in the $J_1-J_2$ triangular model.

Motivated by this question, we use the variational Monte Carlo (VMC) calculations based on
the flux state~\cite{wen1989csl} of fermionic representation to study the $J_1-J'_1-J_2$ triangular Heisenberg model
with additional TRS breaking chiral interactions.
The model Hamiltonian is defined as
%%%%%%%%%%%%%
\begin{eqnarray}\label{ham2}
H&=&J_1\sum_{\langle ij\rangle_{\rm horizontal}} {\bf S}_{i} \cdot {\bf S}_{j} + J'_1\sum_{\langle ij\rangle_{\rm zigzag}} {\bf S}_{i} \cdot {\bf S}_{j} \nonumber\\
&&+J_2\sum_{\langle\langle ij\rangle\rangle} {\bf S}_{i} \cdot {\bf S}_{j}
 + J_{\chi}\sum_{\bigtriangleup/\bigtriangledown} {\bf S}_{i} \cdot ({\bf S}_{j}\times{\bf S}_{k}),
\end{eqnarray}
%%%%%%%%%%%%%
where $J_1$ and $J'_1$ are the horizontal and zigzag NN couplings, respectively (see Fig. \ref{wf}(a)).
We set $J_1=1.0$ as energy scale, and focus on the phase regime with $0.96\le J'_1\le 1.04$
and $0\le J_2 \le 0.15$. The chiral couplings $J_{\chi}$ have the same magnitude in each triangle 
(up triangle $\bigtriangleup$ and down triangle $\bigtriangledown$) as shown in Fig. \ref{wf}(a), 
and the sites $i$, $j$, and $k$ follow the clockwise order in all triangles.
In recent VMC calculations by Zhang, {\it et. al}, some topological features of the spin liquids constructed
based on fermionic flux states have been obtained \cite{Zhang2011,Zhang2012,Zhang2013}.
In particular, the VMC studies
find the CSL in the extended kagom\'{e} model by showing the ground-state degeneracy
and topological Chern number\cite{hu2015csl}. In our calculations, we will follow these techniques.

Through our VMC calculations, we find that while the $120^{\circ}$ antiferromagnetic
order vanishes at a finite chiral coupling $J_{\chi}$ for $J_2 \lesssim 0.08$, the gapless U(1)
DSL in the non-magnetic phase $0.08 \lesssim J_2 \lesssim 0.15$ has the instability towards a CSL
as soon as we turn on the $J_{\chi}$ term. This CSL has a quantized topological Chern number $C=1/2$
and two-fold topological degenerate ground states, which characterize the CSL 
as the $\nu = 1/2$ fractional quantum Hall state. We also study the relation between the chiral order and the lattice 
anisotropy of the $J_1$ coupling in the CSL phase regime. We find the consistent behaviors 
with the DMRG results~\cite{Hu2015} that some spin coupling anisotropy may enhance the robustness of the  CSL.
Our VMC results indicate a positive direction to stabilize a CSL near
the non-magnetic phase in the $J_1-J_2$ triangular Heisenberg model.

\section{VARIATIONAL WAVE FUNCTIONS}
%%%%%%%%%%%%%
\begin{figure}
\begin{center}
\includegraphics[width=\columnwidth]{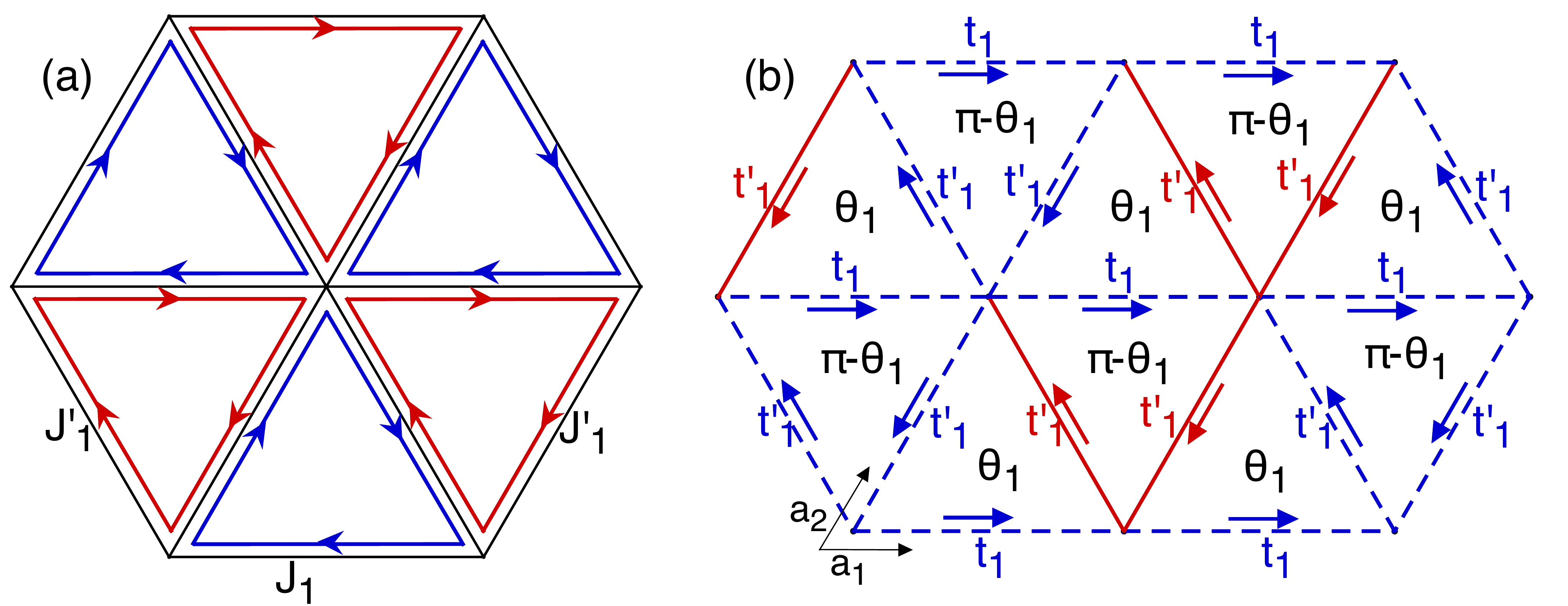}
\end{center}
\caption{ \label{wf}
(Color online) Model Hamiltonian and variational {\it Ansatz}.
(a) In the $J_{1}{-}J'_1{-}J_{2}{-}J_{\chi}$ model Eq.(\ref{ham2}),
we add the same $J_{\chi}$ in the up (blue) and down (red) triangles. 
(b) The variational {\it Ansatz} with the NN hopping $t_{1}$ and $t'_1$ is shown. 
Solid (dashed) lines indicate positive (negative) hoppings, which define the U(1) DSL. 
The phases $\phi_{1}$ and $\phi'_{1}$ are added upon this {\it Ansatz} to obtain a CSL. 
The direction of arrows indicates one possible convention of phases. 
In each up triangle, the flux is $\theta_1=\phi_1+2\phi'_1$; 
in each down triangle, the flux is $\theta_2=\pi-\theta_1$.}
\end{figure}
%%%%%%%%%%%%%

Following one of the novel ways to construct spin liquid states
beyond the mean-field level, we introduce the projected fermionic 
wave functions for our variational calculations~\cite{wen2002}.
In this representation, spin operator ${\bf S}_i$ is expressed using the spinon
operators as ${\bf S}_i = \frac{1}{2} c^{\dagger}_{i,\alpha}{\boldsymbol \sigma}_{\alpha \beta} c_{i,\beta}$,
where ${\boldsymbol \sigma}= (\sigma^x, \sigma^y, \sigma^z)$ is the Pauli matrices
and $c^{\dagger}_{i,\sigma}$ ($c_{i,\sigma}$) creates (annihilates) an electron
with spin $\sigma$ at site $i$. Therefore, the Hamiltonian Eq.~(\ref{ham2}) could
be represented using the fermionic operators,
and the Gutzwiller projector $\mathcal{P}_{G}=\prod_{i}(1-n_{i\uparrow}n_{i\downarrow})$ 
is introduced to enforce no double occupation on each site.
For the variational calculations, we define the variational wave function as
%%%%%%%%%%%%%
\begin{equation}\label{eq:vwf}
|\Psi_{v}\rangle=\mathcal{J}_{s}\mathcal{P}_{G}|\Psi_{0}\rangle,
\end{equation}
%%%%%%%%%%%%%
where $\mathcal{J}_{s}=\exp(1/2 \sum_{ij}v_{ij}S^{z}_{i}S^{z}_{j})$ is the spin Jastrow factor
describing magnetic orders. The variational parameters $v_{ij}$ depend on the distance between sites $i$ and $j$.
$|\Psi_{0}\rangle$ is an uncorrelated ground state 
of mean-field Hamiltonian. In the previous VMC calculations of the $J_1-J_2$ Heisenberg
model~\cite{QiY2015,Iqbal2016}, the $Z_2$ spin liquids have the higher energy than 
the gapless U(1) DSL in the intermediate $J_2$ regime ($0.08 \lesssim J_2 \lesssim 0.16$);
thus, we consider the mean-field Hamiltonian only with the NN hopping term consistent with the DSL,
%%%%%%%%%%%%%
\begin{equation}\label{eq:SL}
{\cal H}_{\rm MF} = \sum_{\langle i,j\rangle,\sigma} t_{ij}c^{\dag}_{i,\sigma}c_{j,\sigma} + h.c.
\end{equation}
%%%%%%%%%%%%%
As shown in Fig.~\ref{wf}(b), the solid (dashed) bonds denote the positive (negative) signs of $t_{ij}$,
which define a magnetic flux $\Phi=0$ crossing up triangles 
and $\Phi=\pi$ crossing down triangles (or opposite)\cite{QiY2015,Lu2015,Iqbal2016}.
For the NN hopping, $t_{ij}=t_1$.
Thus, the unit cell is doubled in this DSL. In our study considering
bond anisotropy and CSL, we allow the anisotropy of the NN hoping $t_{ij}$ and $t'_{ij}$
with both real and imaginary parts, i.e.,
$t_{ij} = |t_{ij}| e^{i\phi_{ij}}$. In Fig.~\ref{wf}(b), we show the {\it Ansatz} of the 
variational wave function \cite{Samuel}. Since the requirement of $t^{*}_{ij} = t_{ji}$,
we define the orientation of the hoping terms in this way:
for the hopping from $j$ to $i$, $t_{ij}$ ($t^{*}_{ij}$) has the direction (opposite direction) along
the arrow shown in Fig.~\ref{wf}(b).
Here, we choose the definition that the up triangles have the fluxes $\theta_1=\phi_{1}+2\phi'_{1}$,
and the down triangles have the fluxes $\theta_2=\pi-\theta_{1}$.
Such a state can be denoted as $[\theta_1, \pi-\theta_1]$.
Thus, using this symbol, the U(1) DSL has the fluxes $[0,\pi]$,
and the wave functions with non-zero $\theta_1$ describe the states
with spin chirality\cite{Wen1989}. In Ref.~\onlinecite{Iqbal2016}, 
the U(1) DSL has competitive variational energy in the 
non-magnetic phase of the $J_1-J_2$ model.

We will also consider the effect of $J_{\chi}$ to the $120^{\circ}$ N\'{e}el order
for $J_2 \lesssim 0.08$. In this case, we define the magnetic states as
%%%%%%%%%%%%%
\begin{equation}\label{eq:AF}
{\cal H}_{\rm MAG} = \sum_{\langle i,j\rangle,\sigma} (t_{ij}c^{\dag}_{i,\sigma}c_{j,\sigma} + h.c.) 
		+ h\sum_{i}{\bf M}_{i} \cdot {\bf S}_{i},
\end{equation}
%%%%%%%%%%%%%
where the magnetic order is described by variational parameter $h$ and unit vector ${\bf M}_{i}$.
The magnetic long-range order is directly related to a non-zero $h$. 
For describing the $120^{\circ}$ N\'{e}el state, we set
${\bf M}_{i}=(\cos({\bf r}_{i}\cdot {\bf q}+\eta_{i}),\sin({\bf r}_{i}\cdot {\bf q}+\eta_{i}),0)$
(${\bf q}$ is the pitch vector and $\eta_{i}$ is the phase shift for the sites within the same unit cell)
with ${\bf q}=(4\pi/3,0)$. For the unit vector ${\bf M}_{i}$ in the $XY$ plane, the spin Jastrow factor 
$\mathcal{J}_{s}=\exp(1/2 \sum_{ij}v_{ij}S^{z}_{i}S^{z}_{j})$ 
correctly describes spin fluctuations around the classical state
in the $XY$ plane \cite{Manousakis1991}.

In this paper we study the competitions among the $120^{\circ}$ N\'{e}el state, 
the gapless U(1) DSL, and the gapped CSL. We perform variational calculations 
at half filling on toric clusters with $L\times L$ sites under the periodic/antiperiodic
boundary conditions (PBC/APBC). In order to find the energetically favored state, 
we use the stochastic reconfiguration (SR) optimization method~\cite{Sorella} 
to optimize the variational parameters.

%%%%%%%%%%%%%
\begin{figure}[b!]
\begin{center}
\includegraphics[width=\columnwidth]{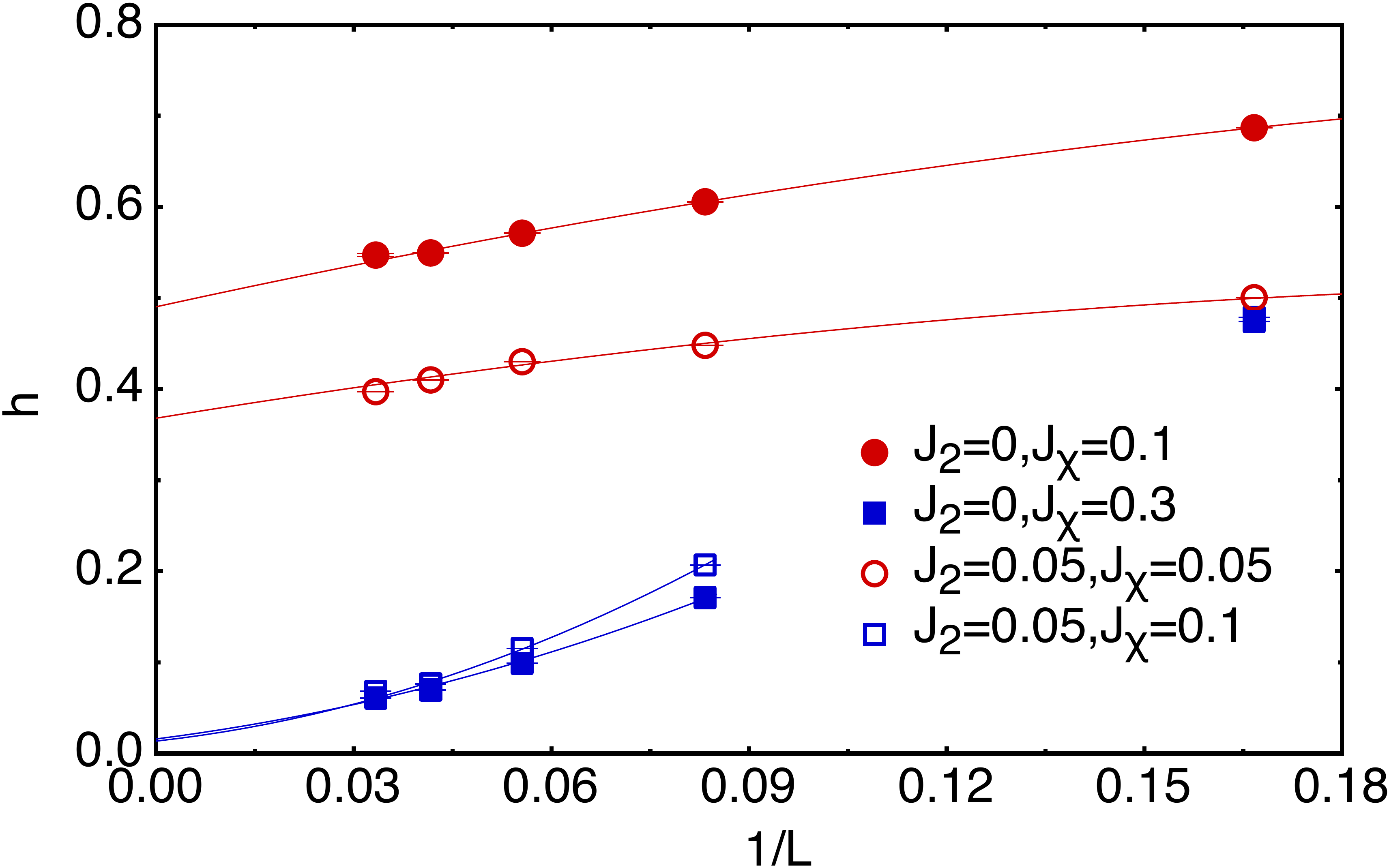}
\end{center}
\caption{ \label{af}
(Color online) Finite-size scaling of the magnetic order variational
parameter $h$ for $J_{2}=0$ and $0.05$ with $J_1=J'_1$. We use the $L\times L$ toric
clusters with PBC at $L = 6,12,18,24,30$. Quadratic fittings are used for all the data.}
\end{figure}
%%%%%%%%%%%%%

\section{Competition between magnetic and chiral orders}
%%%%%%%%%%%%%
\begin{figure}[b!]
\begin{center}
\includegraphics[width=\columnwidth]{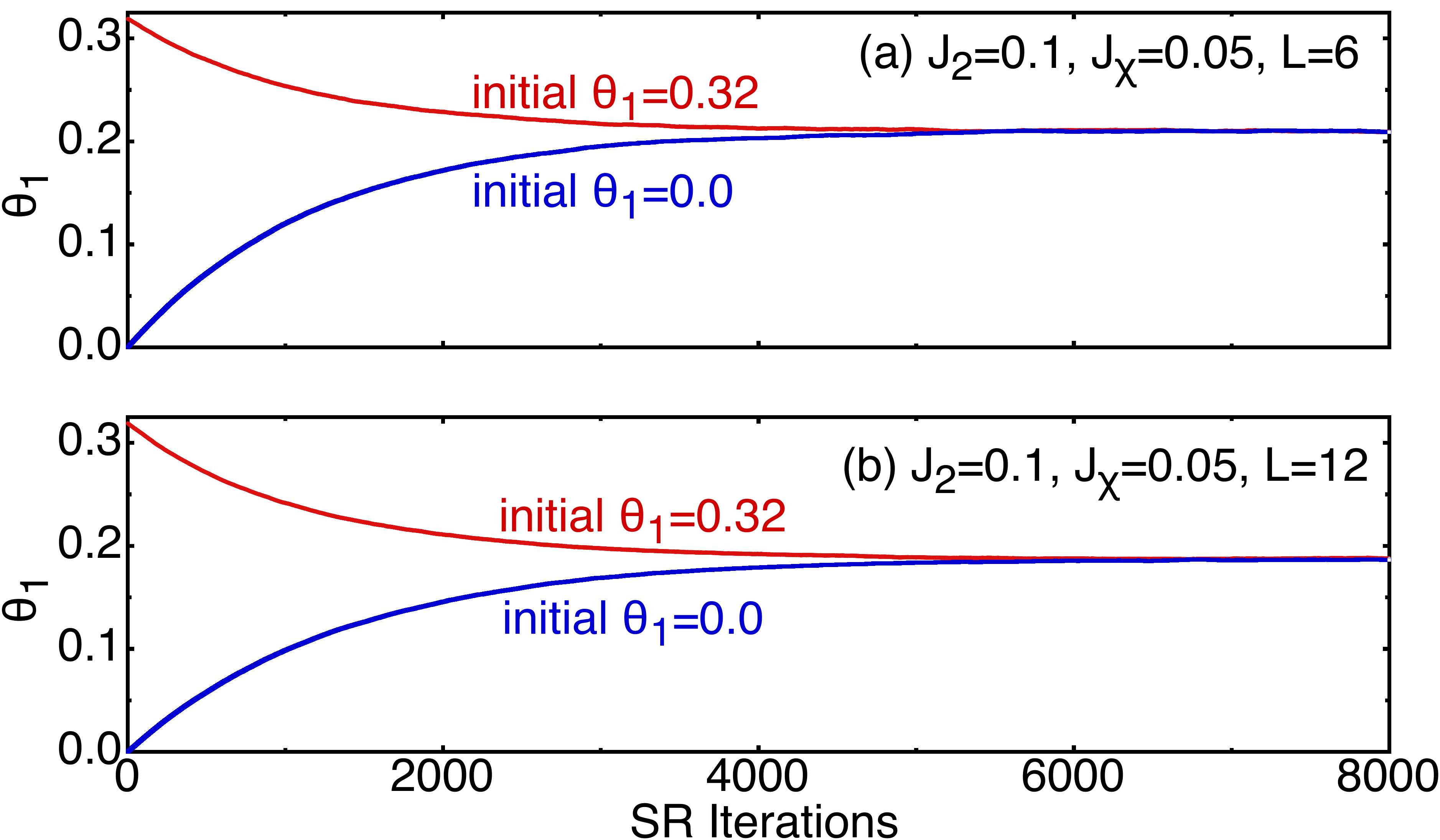}
\end{center}
\caption{ \label{sr}
(Color online) The variational Monte Carlo optimizations of 
the flux $\theta_1=3\arctan(Im(t_1)/Re(t_1))$ are shown 
for $J_{2}=0.1$ and $J_{\chi}=0.05$ on $L=6$ (a) and $12$ (b) clusters. 
Different initial values $\theta_1=0.32$ ($Im(t_1)=0.107$) and $0$ are chosen.}
\end{figure}
%%%%%%%%%%%%%
First of all, we study the competition between the magnetic
and chiral orders for $J_2 \lesssim 0.08$ with $J_1=J'_1$. When $J_{\chi}=0$, the system has the
$120^{\circ}$ N\'{e}el order. When $J_{\chi}$ is much larger than $J_1$ coupling,
the classical spin analyses show that the system would become a non-coplanar
tetrahedral state with four sublattices, where the spins of four sublattices point 
toward the corners of a tetrahedron\cite{messio2011}.
Therefore, with growing $J_\chi$, we expect the system to transit either directly
from the $120^{\circ}$ N\'{e}el order to the tetrahedral phase, or through an intermediate phase.
Interestingly, the CSL discovered in the kagom\'{e} model emerges
between a $120^{\circ}$ N\'{e}el phase and the non-coplanar cuboc phase\cite{Ssgong2015}.
In our present studies, we do not include the variational wave function
of the tetrahedral state. Thus, we only consider the vanishing of the
$120^{\circ}$ N\'{e}el order with the increase of  $J_{\chi}$ (we expect further studies using unbiased methods to investigate
the phase transition between the $120^{\circ}$ N\'{e}el and the tetrahedral
phases in the future work).

In our variational calculations for $J_2 \lesssim 0.08$, we start from the
wave function Eq.~(\ref{eq:AF}) and optimize the parameter $h$ and Jastrow
factor $v_{ij}$. $h = 0$ describes the vanished N\'{e}el order.
We study the lattice with $L=6,12,18,24,30$,
where the $120^{\circ}$ N\'{e}el order is not frustrated by boundary conditions.
In Fig.~\ref{af}, we show the variational parameter $h$ of magnetic order for $J_2=0$ and $0.05$ with $J_{\chi}$ on different clusters.
For both $J_2$ couplings with small $J_{\chi}$, the variational 
parameter $h$ decreases quite slowly with increasing system sizes
and smoothly extrapolates to a finite value in the thermodynamic limit.
When $J_{\chi}$ is large enough ($J_{\chi}\gtrsim0.1$),
%{\bf around what values}
the magnetic order parameter $h$
decreases sharply and scales to vanishing when $L\rightarrow \infty$.
Our results clearly indicate that there is a phase transition with
vanished $120^{\circ}$ N\'{e}el order at a finite $J_{\chi}$.

\section{Chiral spin liquid emerging near the gapless Dirac spin liquid}

In this section we study the possible CSL near the gapless U(1) DSL.
We start from the non-magnetic variational wave function Eq.~(\ref{eq:SL})
without magnetic term ($h=0$) and spin Jastrow factor ($v_{ij}=0$). Thus, 
the variational parameters are the imaginary part of $t_{1}$ and both real and imaginary parts of $t'_{1}$. 
We will focus our studies for $J_2 = 0.1$.

\subsection{Isotropic system with $J_1 = J'_1$}

\subsubsection{Optimization and measurement of local order parameters}

For the isotropic system with $J_1 = J'_1$, we have $t_{1} = t'_{1}$,
and the only variational parameter is the imaginary part of the NN hopping. 
Before discussing the results, we demonstrate the good convergence of our calculations.
In Fig.~\ref{sr}, we show the optimization of $\theta_1$ for $J_2=0.1, J_{\chi}=0.05$ on the
$6\times 6$ and $12\times 12$ clusters. We obtain the converged $\theta_1$
after optimization, which are found to be independent of initial values.

\begin{figure}[b!]
\begin{center}
\includegraphics[width=\columnwidth]{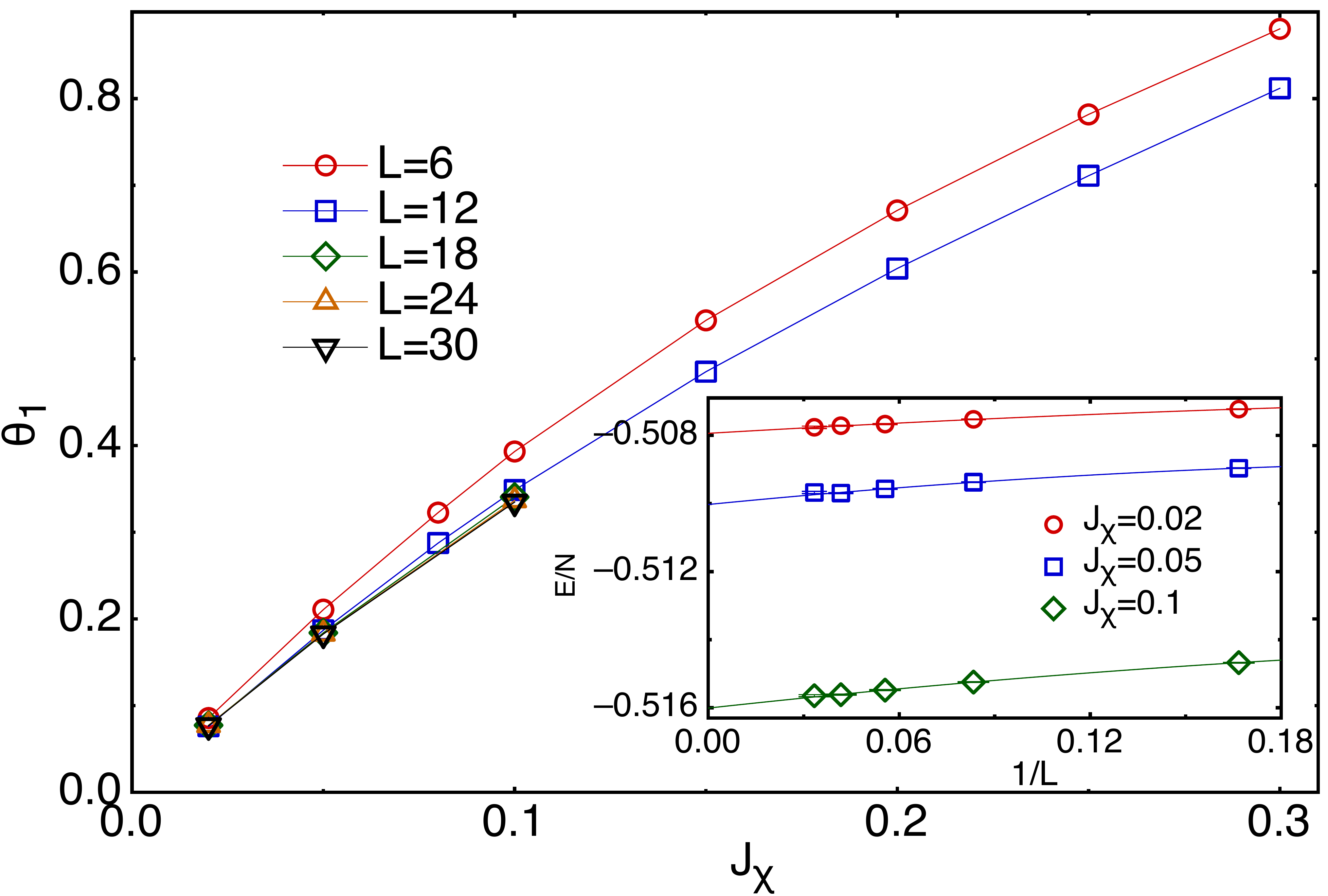}
\end{center}
\caption{ \label{energy}
(Color online) The optimized flux $\theta_1$ in the variational wave function 
at $J_2=0.1$ with different $J_{\chi}$ on $L=6$, $12$, $18$, $24$, and $30$ clusters. 
The inset is the finite size scaling for the ground state energy at $J_{\chi}=0.02$, $0.05$, and $0.1$.}
\end{figure}

To study the CSL, we optimize the variational wave function for different
$J_{\chi}$ on different system size with $L$ up to $L = 30$. On the $L=6$
and $L=12$ clusters, we study the system with $J_{\chi}$ up to $0.3$. As
shown in Fig.~\ref{energy}, the optimized flux $\theta_1$ 
increases with the growing $J_{\chi}$. For $J_{\chi}=0.02,0.05,0.1$, we
study the larger clusters, which give the optimized $\theta_1$ that almost
do not change with increasing system size. In the inset of Fig.~\ref{energy}, 
we also show the finite-size scaling of the ground-state energy for small 
$J_{\chi}$, which support the good convergence of the calculations with system size.
%%%%%%%%%%%%%
%%%%%%%%%%%%%

\begin{figure}
\begin{center}
\includegraphics[width=\columnwidth]{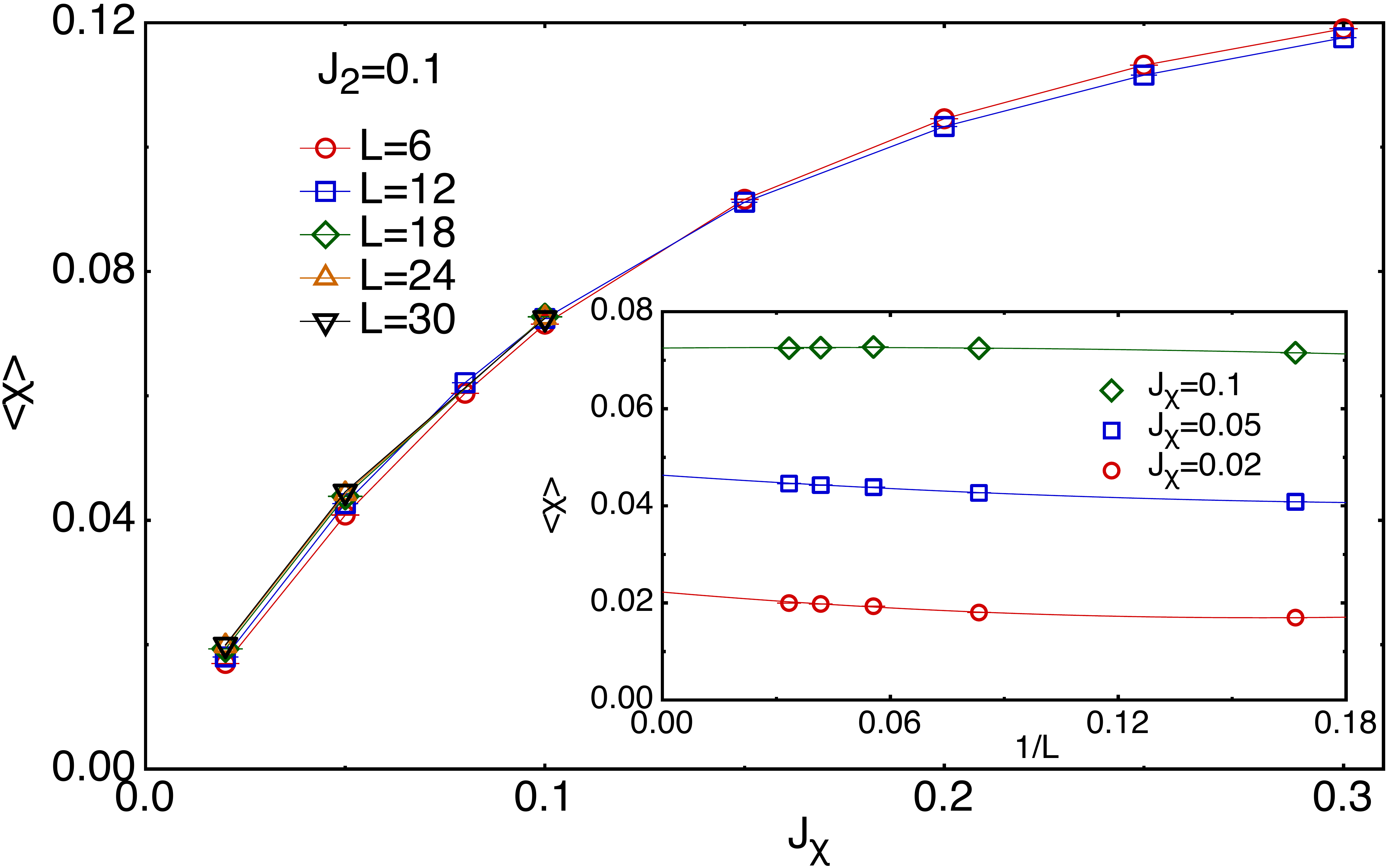}
\end{center}
\caption{ \label{chiral}
(Color online) The chiral order parameter $\langle \chi\rangle$ of the optimized wave functions.
For the up and down triangles, we obtain the same value of $\langle \chi\rangle$ within the error bar.
The inset is the finite size scaling for the chirality $\langle \chi\rangle$ at $J_{\chi}=0.02$, $0.05$, and $0.1$.}
\end{figure}

With the optimized finite variational parameter $\theta_1$, we expect non-zero
chiral order of the optimized wave function, which can be
measured through the three spins scalar chirality in each triangle as:
%%%%%%%%%%%%%
\begin{equation}\label{defchi}
\langle\chi\rangle=\langle{\bf S}_{1} \cdot ({\bf S}_{2}\times{\bf S}_{3})\rangle.
\end{equation}
%%%%%%%%%%%%%
In our calculations, we find that the chiral order parameter $\langle \chi\rangle$ 
of the up and down triangles are the same within the error bar.
In Fig.~\ref{chiral}, we show $\langle\chi\rangle$ as a function of the chiral coupling $J_{\chi}$ 
on different clusters up to $L=30$, which grows continuously with increasing
$J_{\chi}$. The finite-size scaling in the inset of Fig.~\ref{chiral} clearly demonstrates
the non-zero chirality in the thermodynamic limit.

%%%%%%%%%%%%%
%%%%%%%%%%%%%

\subsubsection{Topological properties}

In order to characterize the non-trivial topological properties of the chiral state,
we calculate the topological Chern number and the ground state degeneracy.

%%%%%%%%%%%%%
\begin{figure*}
\begin{center}
\includegraphics[width=0.9\columnwidth]{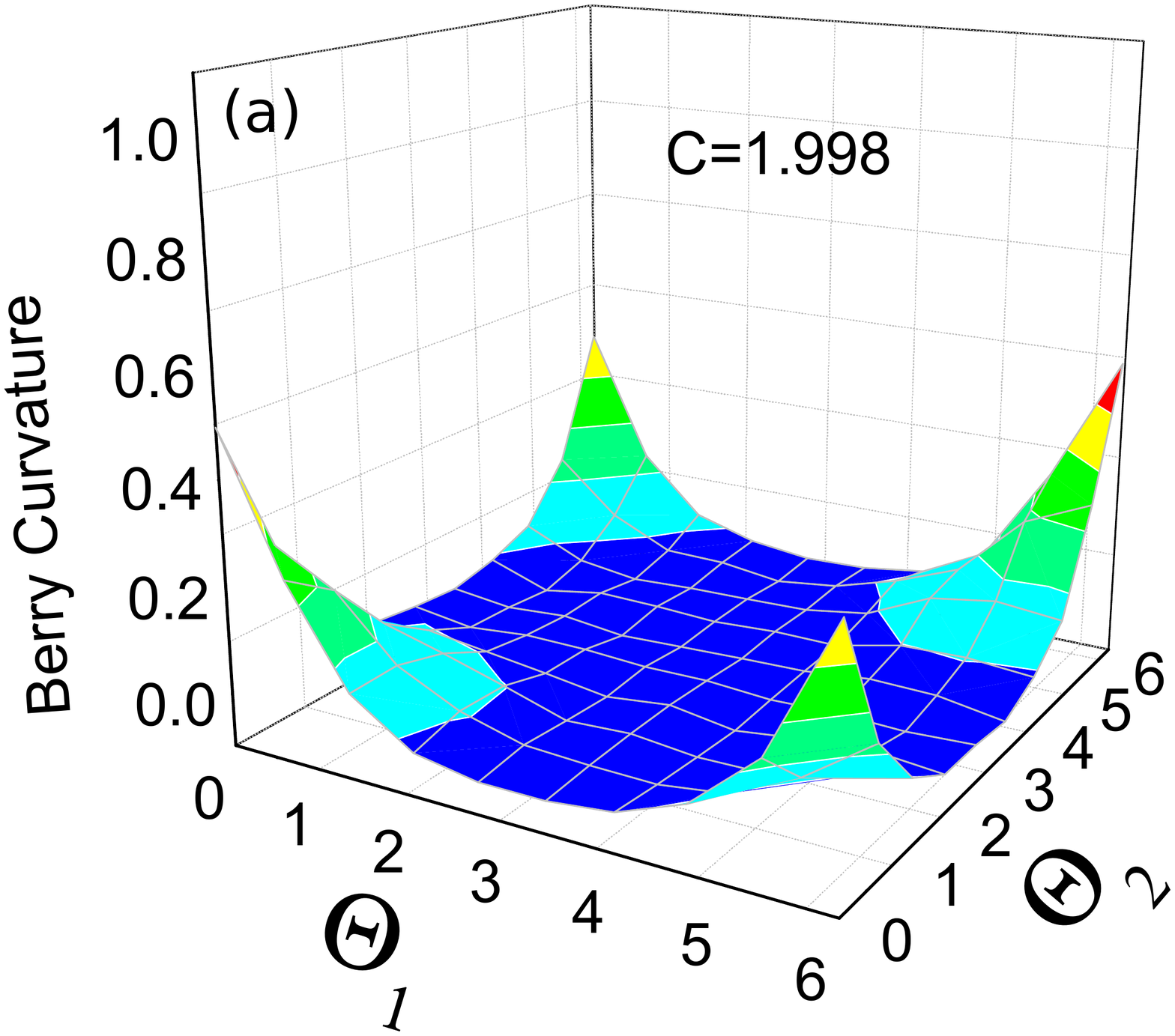}\quad
\includegraphics[width=0.9\columnwidth]{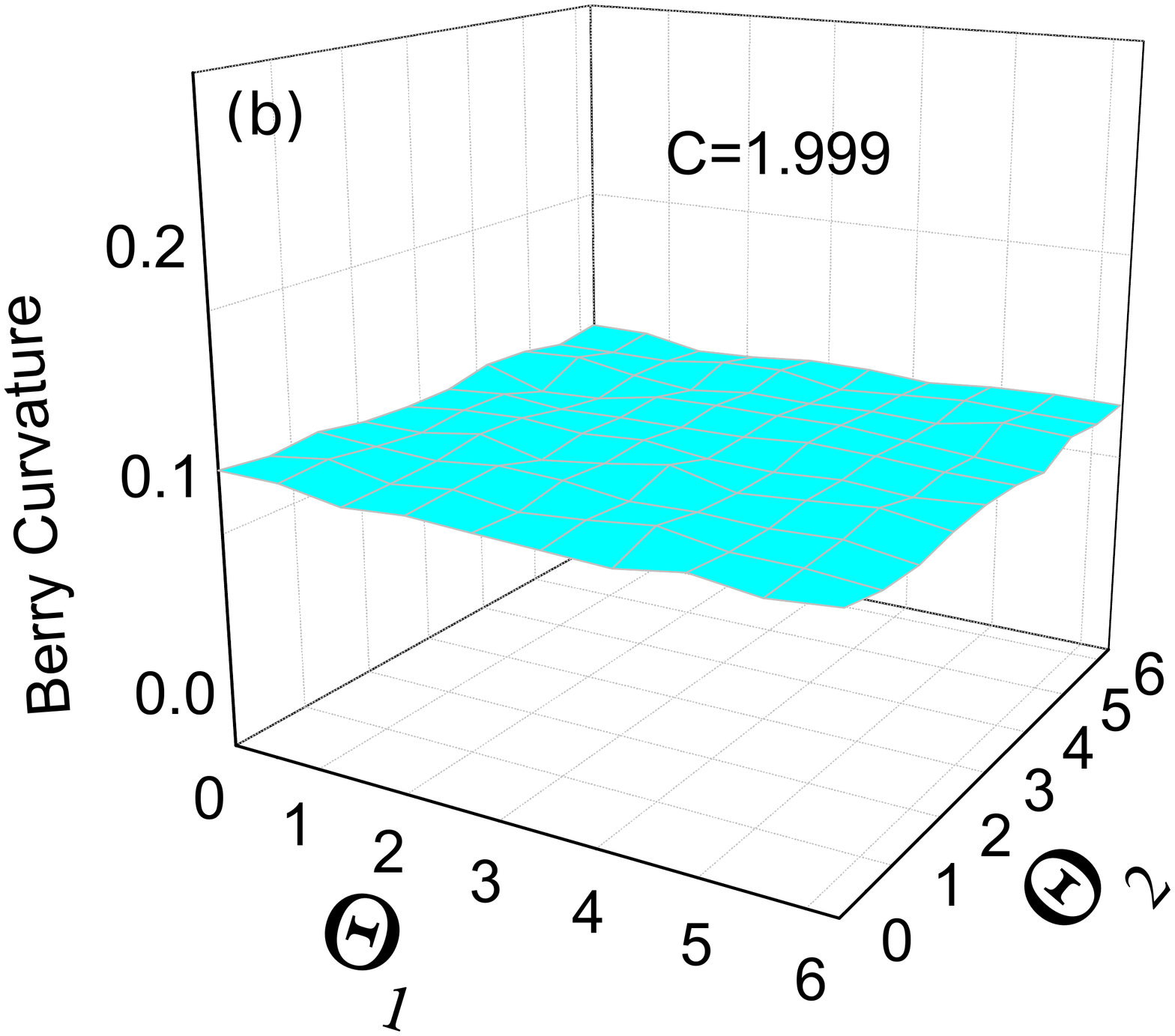}
\end{center}
\caption{ \label{berry}
(Color online) Berry curvature for the states of (a) $J_2=0.1,J_{\chi}=0.05$ 
on the $L=12$ lattice and (b) $[\pi/2,\pi/2]$ on the $L=8$ lattice. For both calculations, 
the Brillouin zone is divided into a mesh with 100 plaquettes. The summation between $0$ and $2\pi$ gives $C=1.998$ (a) and $1.999$ (b).}
\end{figure*}
%%%%%%%%%%%%%
In our calculations, the topological Chern number is computed as 
the integral over the Berry curvature $F(\Theta_{1},\Theta_{2})$ in boundary phase space:~\cite{niu1985,sheng2003,xwan,hafezi}
%%%%%%%%%%%%%
\begin{equation}\label{chern}
C=\frac{1}{2\pi}\int d\Theta_{1}d\Theta_{2}F(\Theta_{1},\Theta_{2}),
\end{equation}
%%%%%%%%%%%%%
where $0\leq\Theta_{k}\leq2\pi$ ($k=1,2$) are twist boundary phases for the torus systems.
To obtain this integral, we uniformly divide the boundary phase space
into $M$ plaquettes ($M$ is chosen up to $100$).
The Berry curvature defined for each plaquette $l$ is calculated as 
$F_{l}=arg\prod_{i=1}^{4}\langle \Psi^{l_{i+1}}_{V}|\Psi^{l_{i}}_{V}\rangle$
($l=1, \dots, M$). The label $i$ ($i=1,2,3,4$) denotes the four corners 
of the $l$-$th$ plaquette, where the periodic condition requires $\Psi^{l_5}_{V}=\Psi^{l_1}_{V}$. 
The wave function $|\Psi^{l}_{V}\rangle$ is the optimized wave function
of the mean-field Hamiltonian with twisted boundary conditions,
which have the opposite requirements for the spin up and spin down partons,
namely $c_{j+L_{k}\uparrow}=c_{j\uparrow}e^{i\Theta_{k}}$ and 
$c_{j+L_{k}\downarrow}=c_{j\downarrow}e^{-i\Theta_{k}}$ ($k=1$ and $2$, and $L_1=L_2=L$ in our calculations).
The overlap for the Berry curvature 
$\langle \Psi^{l_{i+1}}_{V}|\Psi^{l_{i}}_{V}\rangle=\sum_{x}P(x)\frac{\langle x|\Psi^{l_{i}}_{V}\rangle}{\langle x|\Psi^{l_{i+1}}_{V}\rangle}$
is calculated by Monte Carlo method according to the weight 
$P(x)=\frac{|\langle x|\Psi^{l_{i+1}}_{V}\rangle|^{2}}{\sum_{x}|\langle x|\Psi^{l_{i+1}}_{V}\rangle|^{2}}$. 
We obtain the Berry curvatures as shown in Fig.~\ref{berry}.
Here, we consider two wave functions. One is the optimized state at 
$J_2=0.1, J_{\chi}=0.05$ (flux is obtained as $\theta_1\approx0.18$),
and the other one for comparison is the state with fluxes $[\pi/2,\pi/2]$.
For both states, we do the integration of the Berry curvature from $0$ to $2\pi$.
We must emphasize that the integration from $0$ to $2\pi$ 
for the operators of two partons (with spin up and spin down) 
includes two periods of phases for the spin operators, 
so the final results of the Chern number must be divided by $4$ for the spin system. 
In our calculations, the integrations between $0$ and $2\pi$ for both 
states give the results $2$ with high accuracy, which leads to a Chern number $C=1/2$.

In the variational approach with Gutzwiller projected parton construction,
the degeneracy of the wave function is consistent with number of the linear independence 
states of the fermionic variational wave functions according to the
SU(2) Chern-Simons theory.~\cite{Zhang2011,Zhang2012,Zhang2013}
The idea~\cite{Zhang2011,Zhang2012,Zhang2013} is that through changing 
the boundary conditions of the mean field Hamiltonian to either periodic
or antiperiodic in $\vec{a}_{1}$ and $\vec{a}_{2}$ directions (see Fig.~\ref{wf}(b)), 
we can obtain four projected states denoted as $|\psi_{1},\psi_{2}\rangle$.
We label $\psi_{i}=0$ for periodic boundary condition and $\pi$ for antiperiodic boundary condition ($i=1,2$),
i.e., these four projected states are $\{|0,0\rangle, |0,\pi\rangle, |\pi,0\rangle, |\pi,\pi\rangle\}$.
Then we can calculate the overlaps between any two of the four states to 
obtain the overlap matrix~\cite{Zhang2011,Zhang2012,Zhang2013}, and
the number of the nonzero eigenvalues of this overlap matrix gives the 
number of the linearly independent states.
Based on the experience of the similar calculations on kagom\'{e} antiferromagnet~\cite{hu2015csl},
we should choose a state with a big mean-field band gap
to suppress the strong finite-size effects on small clusters.
Thus, we calculate the overlap matrix on the $8\times 8$ cluster
for the state with flux $\theta_1=\pi/2$ (this state has a big mean-field band gap $4.1$),
and obtain the overlap matrix  ${\cal O}$ as

\begin{eqnarray}\label{biggap}
{\cal O} &=&\left(\begin{array}{cccc}
\langle 0,0|0,0\rangle & \langle 0,0|0,\pi\rangle & \langle 0,0|\pi,0\rangle & \langle 0,0|\pi,\pi\rangle \\ 
\langle 0,\pi|0,0\rangle & \langle 0,\pi|0,\pi\rangle & \langle 0,\pi|\pi,0\rangle & \langle 0,\pi|\pi,\pi\rangle \\
\langle \pi,0|0,0\rangle & \langle \pi,0|0,\pi\rangle & \langle \pi,0|\pi,0\rangle & \langle \pi,0|\pi,\pi\rangle \\
\langle \pi,\pi|0,0\rangle & \langle \pi,\pi|0,\pi\rangle & \langle \pi,\pi|\pi,0\rangle & \langle \pi,\pi|\pi,\pi\rangle
\end{array}\right) \nonumber\\
&\approx&\left(\begin{array}{cccc}
1 & 0.57 & 0.57 & 0.58 \\ 
0.57 &1 & 0.58e^{-i1.59} & 0.58e^{i1.59} \\
0.57 & 0.58e^{i1.59} &1 & 0.58e^{-i1.59} \\
0.57 & 0.58e^{-i1.59} & 0.58e^{i1.58} & 1
\end{array}\right) \nonumber\\
&\approx&\left(\begin{array}{cccc}
1                  & \frac{1}{\sqrt{3}}  & \frac{1}{\sqrt{3}}  & \frac{1}{\sqrt{3}}  \\ 
\frac{1}{\sqrt{3}} & 1                   & -\frac{i}{\sqrt{3}} & \frac{i}{\sqrt{3}}  \\
\frac{1}{\sqrt{3}} & \frac{i}{\sqrt{3}}  & 1                   & -\frac{i}{\sqrt{3}} \\
\frac{1}{\sqrt{3}} & -\frac{i}{\sqrt{3}} & \frac{i}{\sqrt{3}}  & 1
\end{array}\right).
\end{eqnarray}
In this calculation, we fix the global phases in such 
a way that all the overlaps with $|0,0\rangle$ are set to real.
The number of the independent ground states can be found by 
diagonalizing the overlap matrix, i.e., ${\cal O}=U^{\dag} \Lambda U$. 
We find that only two eigenvalues are non-zero, which indicates
that only two eigenvectors are linearly independent.
This fact implies that the ground-state degeneracy is two-fold.
Our calculations of topological Chern number and ground state degeneracy 
consistently suggest that this chiral state is the $\nu=1/2$ Laughlin state.

\subsection{Anisotropic system with $J_1 \ne J'_1$}

%%%%%%%%%%%%%
\begin{figure}[b!]
\begin{center}
\includegraphics[width=\columnwidth]{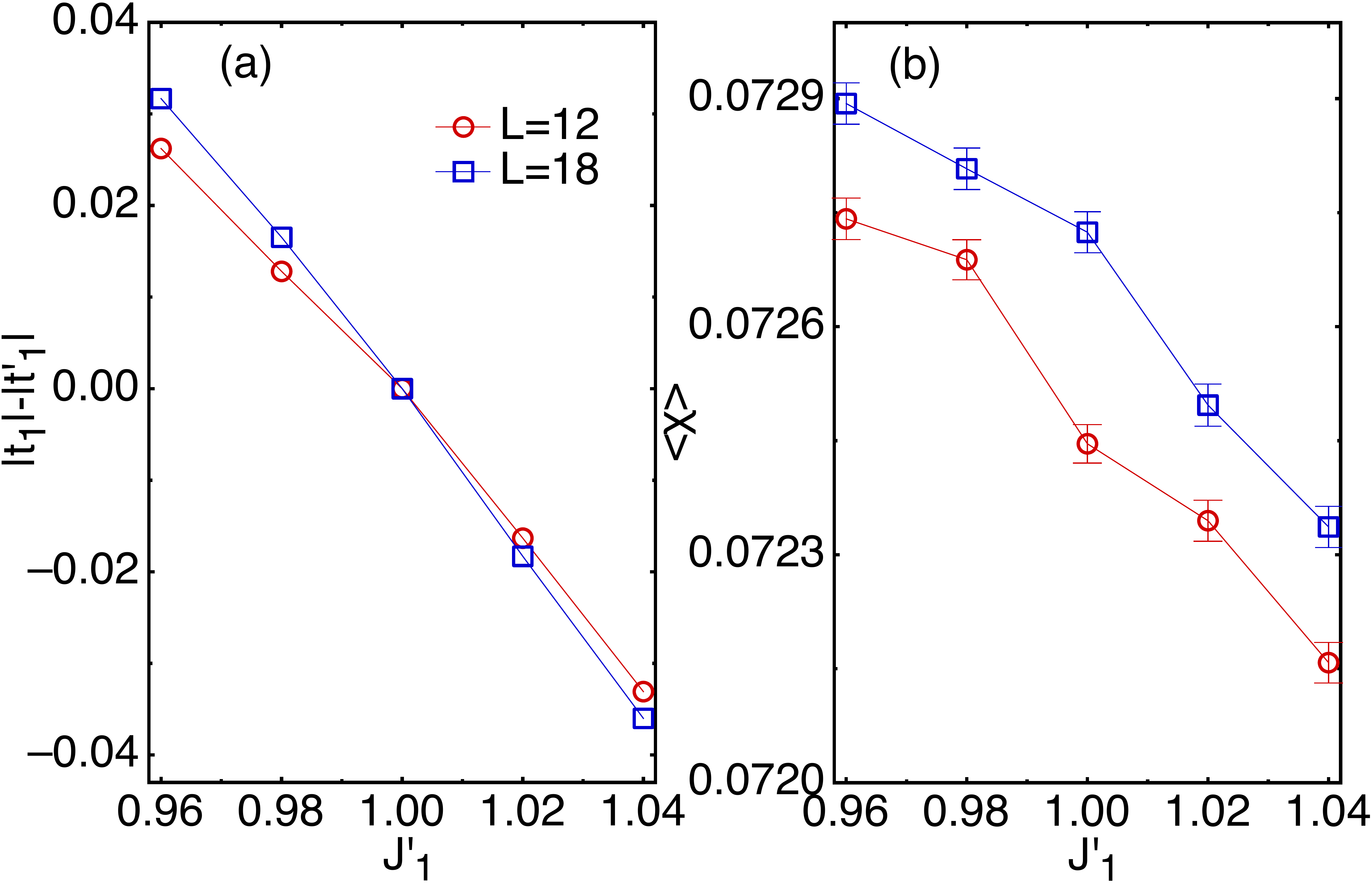}
\end{center}
\caption{ \label{jp}
(Color online). Variational results for the anisotropic system with $J_1 \neq J'_1$.
(a) The difference of the hoppings along horizontal and zigzag directions 
for $0.96\leq J'_1\leq 1.04$ on $L=12$ and $18$ clusters at $J_2=0.1$ and $J_{\chi}=0.1$. 
(b) The chiral order parameter $\langle \chi \rangle$ as function of anisotropy $J'_1$ on $L=12$ and $18$ clusters at $J_2=0.1$ and $J_{\chi}=0.1$.}
\end{figure}
%%%%%%%%%%%%%

In the DMRG studies on the $J_1-J_2$ triangular model, a weak chiral order is found
on the finite-size system in the even sector, and by tuning the bond anisotropy
$J_1$ and $J'_1$ ($J_1$ and $J'_1$ are along the vertical and the zigzag directions, respectively)
the chiral order seems to enhance with $J_1 - J'_1$ for $0.96 \lesssim J'_1/J_1 \lesssim 1.04$\cite{Hu2015}.
The bond anisotropy and chiral order appear to have interesting competition.
In this part, we introduce the bond spatial anisotropy in the $J_1{-}J_2{-}J_{\chi}$ model
to study this competition. We choose $J_{2}=J_{\chi}=0.1$ and
change the anisotropy $J'_1$ from $0.96$ to $1.04$. Correspondingly, 
we use the variational wave function with $t_1\ne t'_1$ (see Fig. \ref{wf}(b)).
Thus, there are three variational parameters (imaginary part of $t_1$, real and imaginary parts of $t'_1$),
including two fluxes that need to be optimized.
Since the optimizations with two fluxes are very time consuming, 
we only did variational calculations on the $L=12$ and $18$ clusters. 

As shown in Fig.~\ref{jp}(a), we find that once $J'_1 \ne J_1$, we obtain
$|t_1| \ne |t'_1|$ after optimization, which indicates that the optimized
wave functions break lattice rotational symmetry. Then, we measure the
spin chirality $\langle\chi\rangle$ for different $J'_1$. Interestingly,
as shown in Fig.~\ref{jp}(b), we find that when $J'_1<J_1$, the chiral 
order is enhanced with increasing anisotropy $|J_1 - J'_1|$; on the contrary
when $J'_1>J_1$, chiral order is suppressed with increasing $|J_1 - J'_1|$.

\section{CONCLUSIONS}

We have studied the spin-$1/2$ antiferromagnetic $J_1-J_2$ Heisenberg model 
with additional chiral coupling $J_{\chi}{\bf S}_i \cdot ({\bf S}_j \times {\bf S}_k)$ on the triangular lattice.
By performing the variational Monte Carlo simulations and considering different variational wave functions,
we find that while the $120^{\circ}$ N\'{e}el order vanishes at a finite $J_{\chi}$,
and the gapless U(1) Dirac spin liquid in the intermediate regime would become a 
chiral spin liquid once $J_{\chi}$ starts to grow. By calculating the topological
Chern number and ground-state degeneracy, we identify this CSL as the $\nu=1/2$ 
Laughlin state. We also consider the relation between the chiral order and the 
spacial anisotropy in the model, and we find that the chiral order can be enhanced 
(suppressed) when the anisotropic parameter $J'_1<J_1$ ($J'_1>J_1$), which is consistent
with the DMRG observation. Our results suggest a new way to stabilize a chiral spin
liquid near the $J_1-J_2$ triangular model. Finally we would like to mention that we 
have not considered all the possible variational states, and it is worth to use unbiased 
numerical simulations such as DMRG to clarify the phase diagram and the properties of the ground states.

%Note added. When we finish this work, we notice a related paper for the classification of the CSL on triangular lattice in Ref.~\onlinecite{Samuel},
%and the CSL {\it Ansatz} in our work is the same as 10b in Table 3 of Ref.~\onlinecite{Samuel}.

\section*{ACKNOWLEDGMENTS}
We acknowledge stimulating discussions with O.~I.~Motrunich, A.~Nevidomskyy and S.~Bieri.
This research  is supported by the National Science Foundation through Grants 
No. DMR-1408560 (W.-J.H., D.N.S.) and PREM DMR-1205734 (S.S.G.) at CSUN, 
NSF Grant No. DMR-1350237 and DMR-1309531 at Rice (W.-J.H.),
and the National High Magnetic Field Laboratory that is supported by NSF DMR-1157490 and the State of Florida (S.S.G.).

%%%%%%%%%%%%%%%%%%%%%%%%%%%%%%%%%%%%%%%%%%%%%%%%
\bibliographystyle{apsrev}
\bibliography{csl_triangle}{}

\end{document}